\documentclass[aps,prl,showpacs,superscriptaddress,twocolumn,raggedfooter,raggedbottom,floatfix]{revtex4-1}   

\usepackage{amssymb}
\usepackage{amsmath}
\usepackage{amssymb}
\usepackage{graphicx}
\usepackage{subfigure}
\usepackage{color}
\usepackage{mathrsfs} 

\usepackage{soul}

\usepackage{hyperref}

\begin{document}

\title{Kink-kink and kink-antikink interactions with long-range tails}
\author{Ivan C.\ Christov}
\affiliation{School of Mechanical Engineering, Purdue University, West Lafayette, IN 47907, USA}
\author{Robert J.\ Decker}
\affiliation{Mathematics Department, University of Hartford, West Hartford, CT 06117, USA} 
\author{A.\ Demirkaya}
\affiliation{Mathematics Department, University of Hartford, West Hartford, CT 06117, USA} 
\author{Vakhid~A.~Gani}
\affiliation{Department of Mathematics, National Research Nuclear University MEPhI (Moscow Engineering Physics Institute), 115409 Moscow, Russia}
\affiliation{Theory Department, National Research Center Kurchatov Institute, Institute for Theoretical and Experimental Physics, 117218 Moscow, Russia}
\author{P.~G.\ Kevrekidis}
\affiliation{Department of Mathematics and Statistics, University of Massachusetts,
Amherst, MA 01003, USA} 
\author{Avinash~Khare} 
\affiliation{Physics Department, Savitribai Phule Pune University, Pune 411007, India} 
\author{Avadh~Saxena} 
\affiliation{Theoretical Division and Center for Nonlinear Studies, Los Alamos National Laboratory, Los Alamos, New Mexico 87545, USA} 

\begin{abstract}
In this Letter, we address the {long-range interaction} between kinks and antikinks, as well as kinks and kinks, in $\varphi^{2n+4}$ field theories for $n>1$. The kink-antikink interaction is generically attractive, while the kink-kink interaction is generically repulsive. We find that the force of interaction decays with the $(\frac{2n}{n-1})$th power of their separation, and we identify the general prefactor for {\it arbitrary} $n$. Importantly, we test the resulting mathematical prediction with detailed numerical simulations of the dynamic field equation, and obtain good agreement between theory and numerics for the cases of $n=2$ ($\varphi^8$ model), $n=3$ ($\varphi^{10}$ model) and $n=4$ ($\varphi^{12}$ model).
\end{abstract}

\maketitle


\paragraph*{Introduction.} The study of field-theoretic models with polynomial potentials has been a topic of wide appeal across a diverse span of theoretical physics areas, including notably cosmology, condensed matter physics and nonlinear dynamics~\cite{vilenkin01,manton_book,eilbeck}. Arguably, the most intensely studied model in this class is the quartic (double well) potential, the so-called $\varphi^4$ model, connected to the phenomenological Ginzburg--Landau theory~\cite{GinzburgLandau,Tinkham}, among numerous other applications~\cite{Bishop80,K75,csw,vach08}. While the $\varphi^4$ model has a time-honored history in its own right~\cite{ourbook}, more recently, higher-order field theories have emerged as models of phase transitions \cite{khare} relevant to material science~\cite{gl,ferro,iso} (see also~\cite[Chap.~11]{ourbook} and \cite{Katsnelson14}), or in quantum mechanical problems (including supersymmetric ones)~\cite{bazeia17}, among others. There, the prototypical example has been the $\varphi^6$ field-theoretic model, which has led to numerous insights and novel possibilities with respect to the spectral properties~\cite{DDKCS} and wave interactions~\cite{phi6}.

Scattering of solitary waves (topological defects or otherwise) is a long-standing topic of active research \cite{amin}, starting from the early works \cite{K75,csw}. Our aim here is to go beyond the ``classical'' models, in a direction that, admittedly, has already seen some significant activity~\cite{lohe,mello,bazeia06-11,Gomes.PRD.2012,khare,GaLeLi,bazeia18,Belendryasova.2019}. One of the particularly appealing aspects of this research program (aside from its potential above-mentioned applications in material science or high-energy physics/quantum mechanics) is that higher-order field theories possess topological defect solutions (kinks) with \emph{power-law} tails, rather than the ``standard'' exponential tails that we are used to in the $\varphi^4$ and the (usual variants of) $\varphi^6$ field theories. The resulting dynamics set by the power-law tails endows topological defects with \emph{long-range} interactions. Recently, a methodology for quantifying such kink-kink and kink-antikink interactions in the $\varphi^8$ model was proposed in \cite{manton}. In our previous work \cite{ivan_longrange}, we showed that there are some deep challenges in even initializing such topological defect configurations numerically. Thus, the initial conditions in a direct numerical simulation of interactions may substantially affect the nature of the observed interactions (cf.\ also earlier works including, e.g.,~\cite{Belendryasova.2019}).

One of the related motivations of our study stems from the expectation that power-law tails may affect the physical properties of a system governed by higher-order field theories. For instance, the dynamics and interaction of domain walls in ferroelastic materials undergoing successive phase transitions \cite{ferro} should affect the elastic properties in unusual ways. Similarly, the response of crystals undergoing isostructural transitions \cite{iso} or the behavior of crystallization of chiral proteins \cite{proteins} should be altered by the associated long-range interactions between domain walls. 

The present effort provides an answer to the question of how kink-kink (K-K) and kink-antikink (K-AK) long-range (power-law) interactions occur in higher-order field theories that exhibit such topological defects. We consider particular (yet highly relevant to this problem) potentials where the highest power of $\varphi$ in the potential is $2n+4$
and we analyze the interaction for arbitrary $n\ge2$. We find that kinks repel each other, while kinks and antikinks generically attract, in both cases with a power law decaying as the $(\frac{2n}{n-1})$th power of their mutual separation. Furthermore, we adapt the recent methodology of~\cite{manton} to the case of arbitrary $n\ge2$, and we identify the prefactor (distinct for K-K and K-AK) of the corresponding power-law interaction force. Equally important, we resolve the ``uncertainty'' of the prefactor indicated in~\cite{manton}. We identify the most accurate asymptotic prefactor and test it against direct numerical simulations to find {\it good agreement} for  $\varphi^8$ ($n=2$), $\varphi^{10}$ ($n=3$) and $\varphi^{12}$ ($n=4$) models, for {\it both} K-K and K-AK interactions. The increasing trend of the deviations between the two, 
as $n$ increases, is also explained.

First, we present our theoretical results. Then, we compare them to direct numerical simulations. Finally, we offer some conclusions and directions for future work. 

\paragraph*{Theoretical Analysis.}
Consider a real scalar field $\varphi(x,t)$ in $(1+1)$ dimensional spacetime, its dynamics set by the Lagrangian density
\begin{equation}\label{eq:largang}
	\mathscr{L}=\frac{1}{2} \left( \frac{\partial\varphi}{\partial t} \right)^2-\frac{1}{2} \left( \frac{\partial\varphi}{\partial x} \right) ^2-V(\varphi).
\end{equation}
The dynamic equation of motion of this field is
\begin{equation}\label{eq:nkg}
	\frac{\partial^2\varphi}{\partial t^2} = \frac{\partial^2\varphi}{\partial x^2} - \frac{\mathrm{d}V}{\mathrm{d}\varphi}.
\end{equation}
The potential $V$ is, specifically, of the form 
{
\begin{equation}\label{potential}
V(\varphi)=\frac{1}{2}(1-\varphi^2)^2\varphi^{2n}. 
\end{equation}
}
This potential has three minima: $\bar{\varphi}_1=-1$, $\bar{\varphi}_2=0$, and $\bar{\varphi}_3=1$. Hence, there are two kinks in this model, $\varphi_{(-1,0)}(x)$ and $\varphi_{(0,1)}(x)$, and two corresponding antikinks,  $\varphi_{(0,-1)}(x)$ and $\varphi_{(1,0)}(x)$.  All of these defects exhibit one power-law and one exponential asymptotic decay to the respective equilibria ($0$ and $\pm 1$) 
as $|x|\to \infty$. We study the interaction force between the kink $\varphi_{(0,1)}(x)$ and the kink $\varphi_{(-1,0)}(x)$. {Their  time-dependent positions are $x=\pm A(t)$, respectively, and their long-range tails overlap. Similarly, for the K-AK interaction, we employ the antikink $\varphi_{(0,-1)}(x)$ and the corresponding mirror kink $\varphi_{(-1,0)}(x)$ located at $x=\pm A(t)$, respectively.

In \cite{ivan_longrange}, the interaction via power-law tail asymptotics was studied numerically for $n=2$ ($\varphi^8$ model). In \cite{manton}, the force between a well-separated kink-kink and kink-antikink was analyzed, again for $n=2$. Our aim here is to generalize the (most sophisticated among the different) approach(es) of the very recent work~\cite{manton}, and to calculate the result for arbitrary $n$. Then, we blend this theoretical analysis with the delicate computational approach from~\cite{ivan_longrange} to fully flesh out the K-K and K-AK long-range interactions in such higher-order field theoretic models involving power-law tails. 

Below, we model the accelerating kink solution of Eq.~\eqref{eq:nkg} by a field  of the form $\varphi(x, t) = \phi(y)$, {where $y= x-A(t)$ and $\phi=\varphi_{(0,1)}$ or $\varphi_{(0,-1)}$}.

\paragraph*{Kink-Kink Interaction.}
Substituting 
the kink profile into Eq.~\eqref{eq:nkg} yields the static equation for $\phi$:
\begin{equation}\label{eq:static_nkg}
\phi'' + a\phi' - \left.\frac{\mathrm{d}V}{\mathrm{d}\varphi}\right|_{\varphi=\phi} = 0,
\end{equation}
where $a=\ddot{A}$ is the acceleration (assumed small) and Lorentz contraction terms ($\propto \dot{A}^2$) have been neglected. Here, $\dot{A}=\mathrm{d}{A}/\mathrm{d}t$, and $\phi' = \mathrm{d}\phi/\mathrm{d}y$. Following~\cite{manton}, the Bogomolny equation $\phi'=(\mathrm{d}W/\mathrm{d}\varphi)|_{\varphi=\phi}$, where $V={(1/2)} (\mathrm{d}W/\mathrm{d}\varphi)^2$, is used to eliminate $\phi'$ from Eq.~\eqref{eq:static_nkg}. Treating $a$ as slowly varying, we define an effective potential $\tilde{V}(\phi) \approx V(\phi) - a W(\phi)$. Then, from the first integral of Eq.~\eqref{eq:static_nkg} (setting the constant of integration in the limit of $y\to\infty$), we obtain
\begin{multline}\label{eq:eq_tail}
\left(\frac{\mathrm{d}\phi}{\mathrm{d} y}\right)^2 = 2\tilde{V}(\phi) + 2a W(1)
\sim \phi^{2n}+\frac{4a}{(n+1)(n+3)}.
\end{multline}
This calculation is asymptotic (dropping $\mathrm{o}(\phi^{2n})$ terms), using the fact that our chosen family of potentials is such that $V(\phi) \sim \frac{1}{2}\phi^{2n}$ as $\phi\to0$. Meanwhile, the second term in Eq.~\eqref{eq:eq_tail} (from the contribution of $W(1)-W(0)$, independently of the normalization of $W$) is effectively proportional to the kink's rest mass $M=2 /[(n+1) (n+3)]$ (for arbitrary $n$).  Requesting (as in~\cite{manton}) that $\phi(y)$ should have the properties that $\phi(-A)= 0$ while $\phi(0)$ diverges, we can re-arrange Eq.~\eqref{eq:eq_tail} into a quadrature:
\begin{equation}
  \int_{0}^{\infty} \frac{\mathrm{d}\phi}{\sqrt{\phi^{2n}+\frac{4a}{(n+1)(n+3)}}}=A.
  \label{start}
\end{equation}
The change of variables $\phi=\left({4a}/[(n+1)(n+3)]\right)^{\frac{1}{2n}} \lambda$ in Eq.~\eqref{start} yields
\begin{equation}
\left[\frac{4a}{(n+1)(n+3)}\right]^{\frac{1-n}{2n}} \int_{0}^{\infty}  \frac{\mathrm{d}\lambda}{\sqrt{\lambda^{2n}+1}}=A.
\label{start-change-var}
\end{equation}
The relevant integral can be computed as
\begin{equation}
\int_{0}^{\infty} \frac{\mathrm{d}\lambda}{\sqrt{\lambda^{2n}+1}}=\frac{\Gamma(\frac{n-1}{2n})\Gamma(\frac{1}{2n})}{2n\sqrt{\pi}},
\end{equation}
yielding the acceleration during the K-K interaction:
\begin{equation}\label{kink-kink}
a=\left[\frac{\Gamma(\frac{n-1}{2n})\Gamma(\frac{1}{2n})}{2n\sqrt{\pi}}\right]^{\frac{2n}{n-1}}\frac{(n+1)(n+3)}{4}A^{\frac{2n}{1-n}}.
\end{equation}

For the $\varphi^8$ model, $n=2$, and Eq.~\eqref{kink-kink} yields $a={44.3139}/{A^4}$. From Newton's second law ($F=Ma$), the force is $F=\frac{2}{15}a={5.9085}/{A^4}$. Similarly, for the $\varphi^{10}$ model, $n=3$, and we get  $a={ 16.5411}/{A^3}$ and $F=\frac{1}{12}a={1.3784}/{A^3}$. Finally, for the $\varphi^{12}$ model, $n=4$, and we get  $a={ 16.1871}/{A^{8/3}}$ and $F=\frac{2}{35}a={0.9250}/{A^{8/3}}$.

\paragraph*{Kink-Antikink Interaction.} 
The calculation in the K-AK case proceeds in the same way with the main difference that now $a=-\ddot{A}$ due to the attraction, in this case, between kink and antikink. From the corresponding version of Eq.~\eqref{start}, we have
\begin{equation}
\int_{\left[\frac{4a}{(n+1)(n+3)}\right]^{1/2n}}^{\infty} \frac{\mathrm{d}\phi}{\sqrt{\phi^{2n}-\frac{4a}{(n+1)(n+3)}}}=A.
\label{start2}
\end{equation}
Notice that, now, the integral must be from the turning point rather than from $0$, related to the sign change of the Bogomolny equation satisfied by the anti-kink.

\clearpage

\begin{figure}[tbp]
\includegraphics[width=0.235\textwidth]{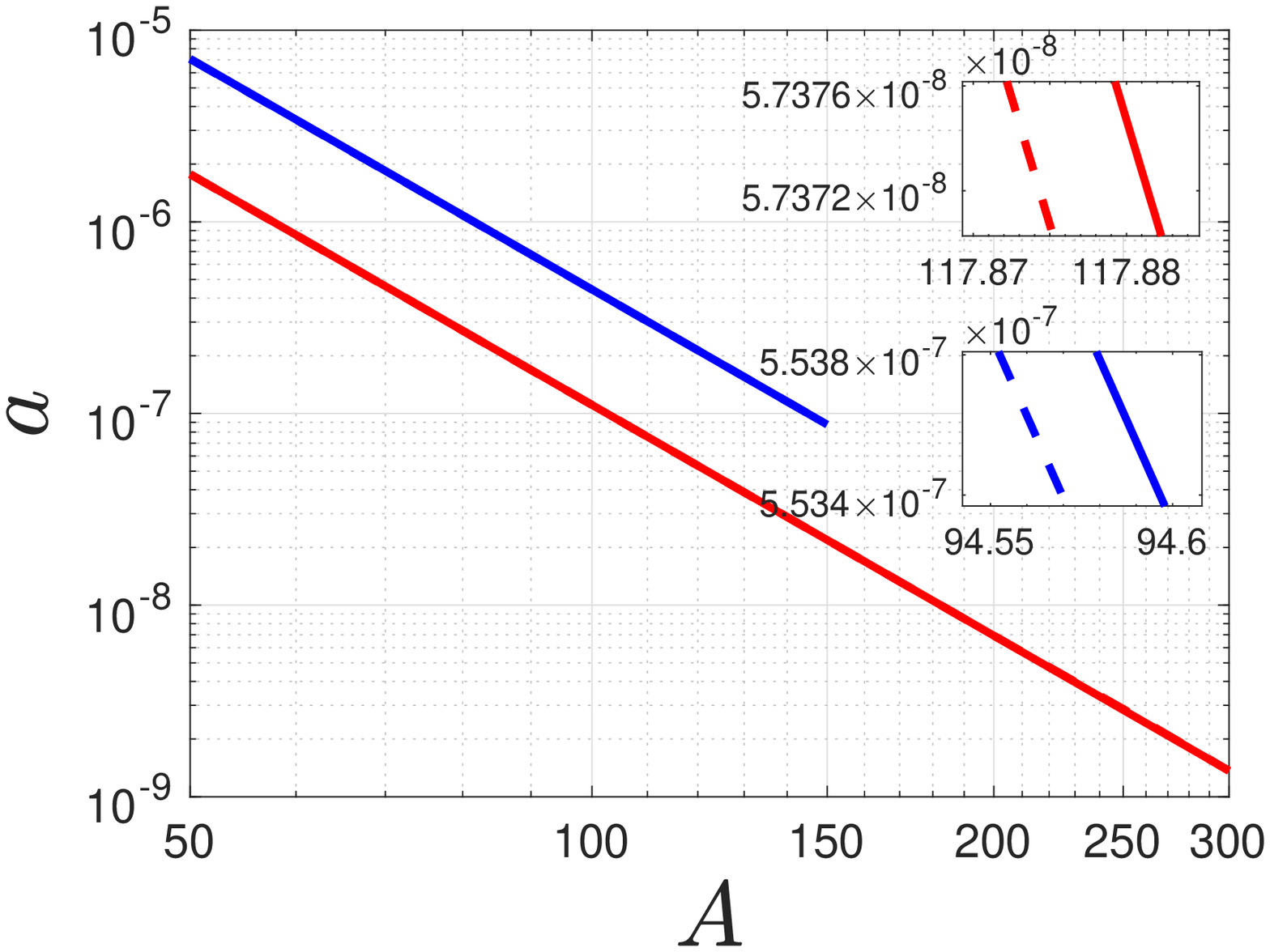}
\includegraphics[width=0.235\textwidth]{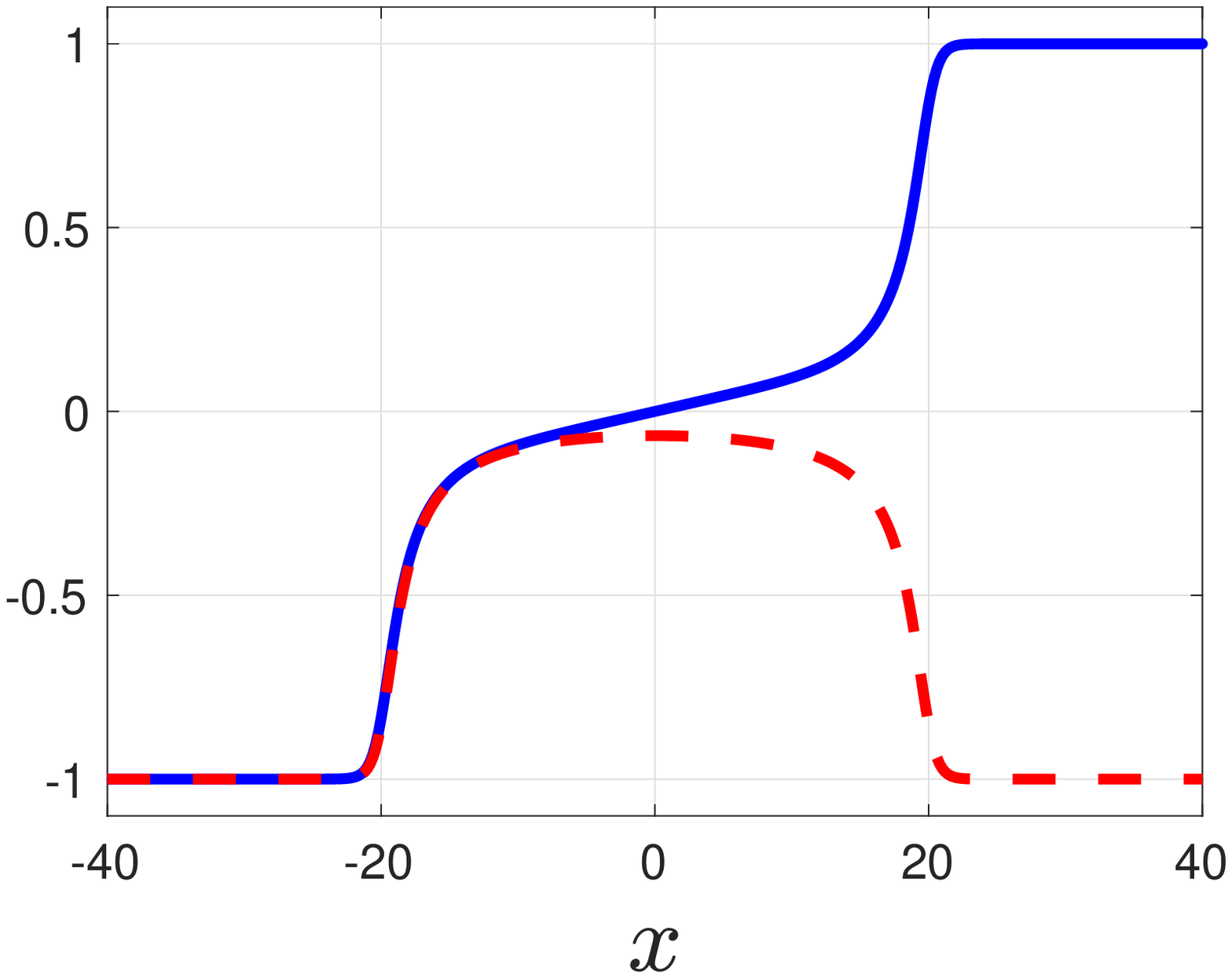}
\includegraphics[width=0.235\textwidth]{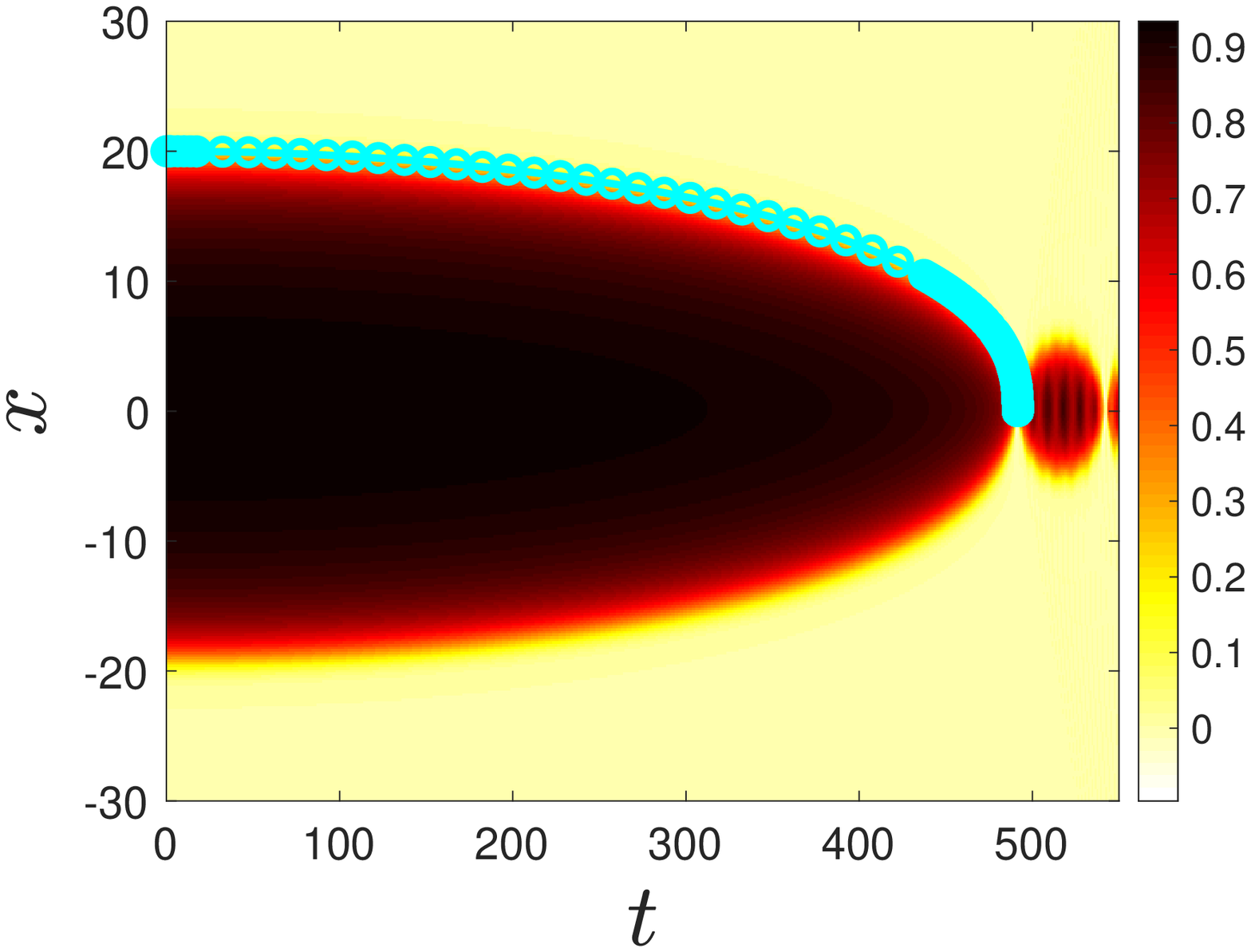}
\includegraphics[width=0.235\textwidth]{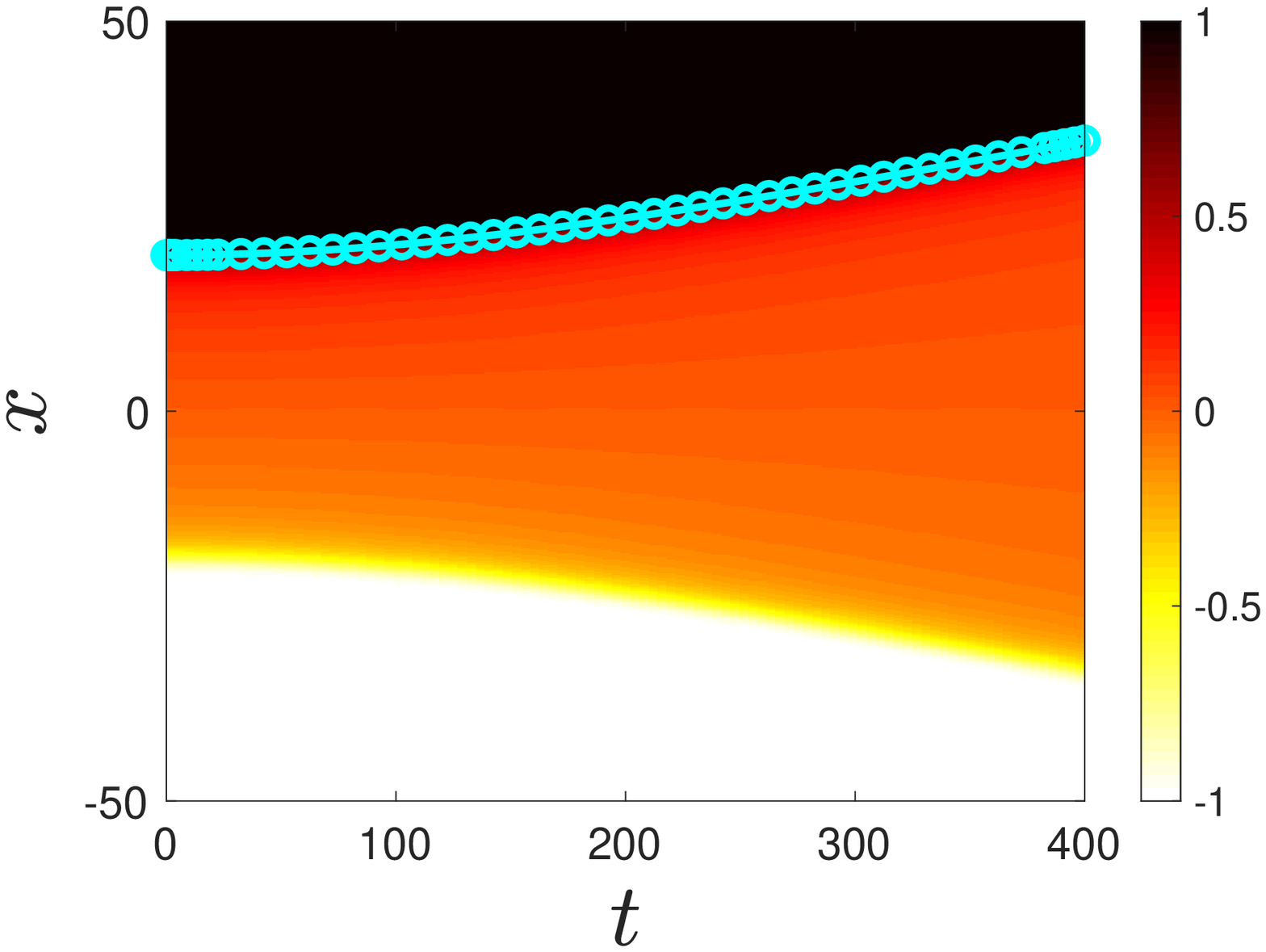}
\caption{Top left is the log-log plot of the left kink acceleration $a$ as a function of $A$ for the kink-kink interaction (blue) and kink-antikink interaction (red) in the $\varphi^8$ model. The dashed lines are computed numerically, and the solid lines are computed analytically. Top right is the plot of the initializers for the kink-kink (K-K) (blue-solid) and kink-antikink (K-AK) (red-dashed) interaction. Bottom left is the space-time contour plot of the K-AK interaction, and the cyan curve with circle symbols is the plot of the solution to the initial value problem (IVP): $\ddot{x}(t)=-11.0785/x^4$, $x(0)=20$, $\dot{x}(0)=0$. Bottom right is space-time contour plot of the K-K interaction,  and the cyan curve with circle symbols is the solution to the IVP: $\ddot{x}(t)=44.3139/x^4$, $x(0)=20$, $\dot{x}(0)=0$.}
\label{fig:powerLaw8}
\end{figure}

Using the same change of variables as above, Eq.~\eqref{start2} becomes
\begin{equation}
\left[\frac{4a}{(n+1)(n+3)}\right]^{\frac{1-n}{2n}} \int_{1}^{\infty}  \frac{\mathrm{d}\lambda}{\sqrt{\lambda^{2n}-1}}=A.
\end{equation}
Once again, the integral can be calculated:
\begin{equation}
\int_{1}^{\infty} \frac{\mathrm{d}\lambda}{\sqrt{\lambda^{2n}-1}}=\frac{-\sqrt{\pi}\,\Gamma(\frac{n-1}{2n})}{\Gamma(-\frac{1}{2n})},
\end{equation}
yielding the acceleration during the K-AK interaction:
\begin{equation}\label{eqahat2a}
a=\left[\frac{-\sqrt{\pi}\,\Gamma(\frac{n-1}{2n})}{\Gamma(-\frac{1}{2n})}\right]^{\frac{2n}{{n-1}}}\frac{(n+1)(n+3)}{4}A^{\frac{2n}{1-n}}.
\end{equation}

An intriguing observation stems from the ratio of Eqs.~\eqref{eqahat2a} and \eqref{kink-kink}. In particular, using the well-known identities
$a \Gamma(a)=\Gamma(a+1)$, and $\Gamma(a) \Gamma(1-a)=\pi/\sin(\pi a)$,
we can express the ratio of the K-AK to K-K forces as
\begin{equation}
  R=\frac{F_\text{K-AK}}{F_\text{K-K}} = - \left[\sin \left(\frac{\pi}{2 n}\right)
    \right]^{\frac{2 n}{n-1}}.
\label{extra1}
\end{equation}
This expression suggests that, contrary to what is known about ``standard'' models such as sine-Gordon or $\varphi^4$ and their exponentially decaying kinks and antikinks~\cite{manton_book,eilbeck,ourbook}, here
the ratio of the K-AK to K-K force is {\it not} equal to $1$, but rather decreases with $n$. Therefore, a fundamental characteristic of long-range-interacting kinks is that this feature becomes more dramatic (with the ratio, in principle, tending to $0$ as $n \rightarrow \infty$), the ``heavier'' the tails. 

For $n=2$ ($\varphi^8$ model), from Eq.~\eqref{eqahat2a}, we get $a={11.0785}/{A^4}$ and $F=-\frac{2}{15}a = -{1.4771}/{A^4}$. Similarly, for $n=3$ ($\varphi^{10}$ model), we get $a={ 2.0676}/{A^3}$ and $F=-\frac{1}{12}a = -{ 0.1723}/{A^3}$. Finally, for $n=4$ ($\varphi^{12}$ model), we get $a={1.2495}/{A^{8/3}}$ and $F=-\frac{2}{35}a=-{0.0714}/{A^{8/3}}$. Armed with these specific predictions for K-K and K-AK interactions, we turn to verification of our general theory via direct numerical simulations.

\paragraph*{Numerical Results.}

\begin{figure}[tbp]
\centering
\includegraphics[width=0.235\textwidth]{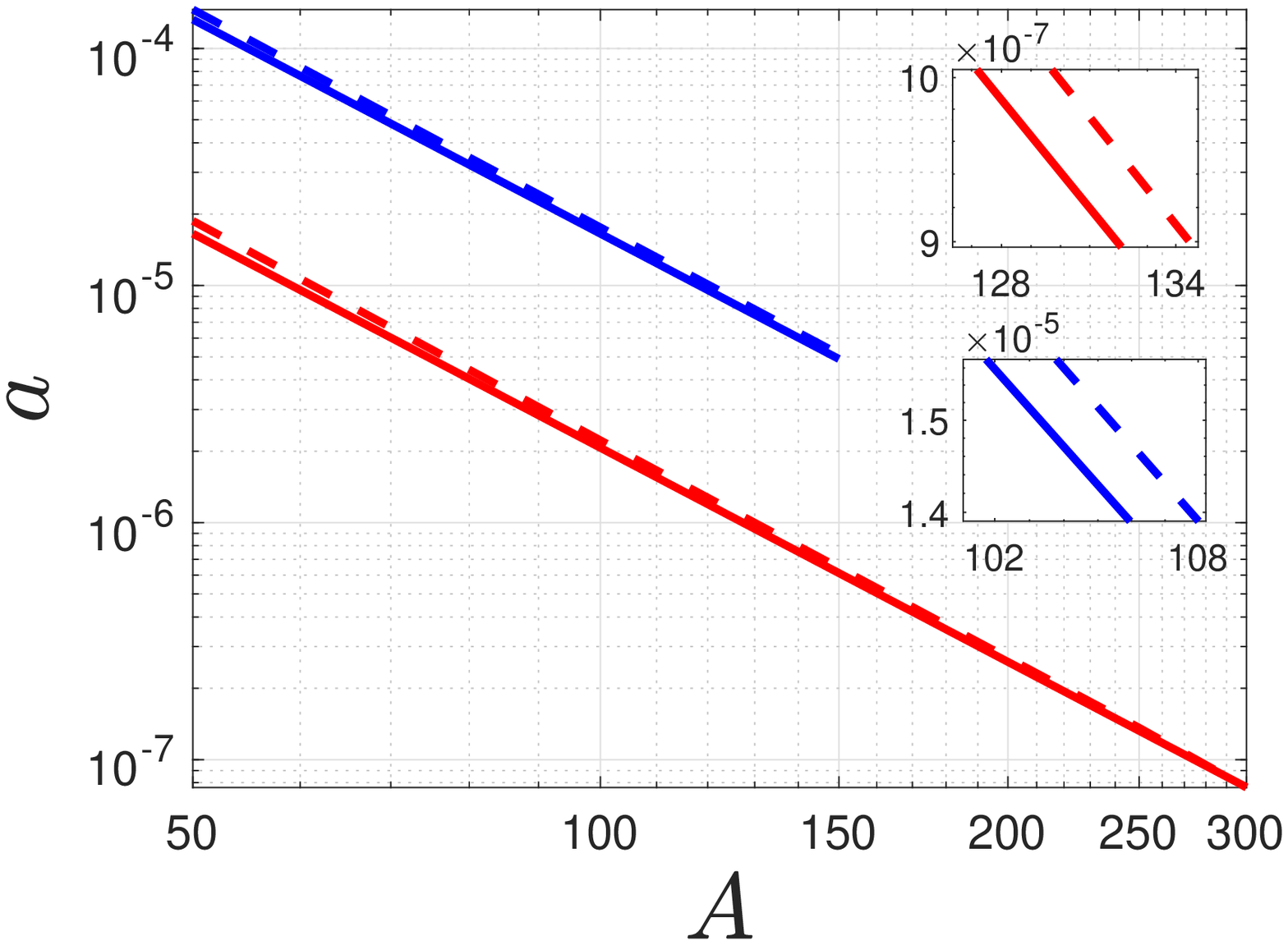}
\includegraphics[width=0.235\textwidth]{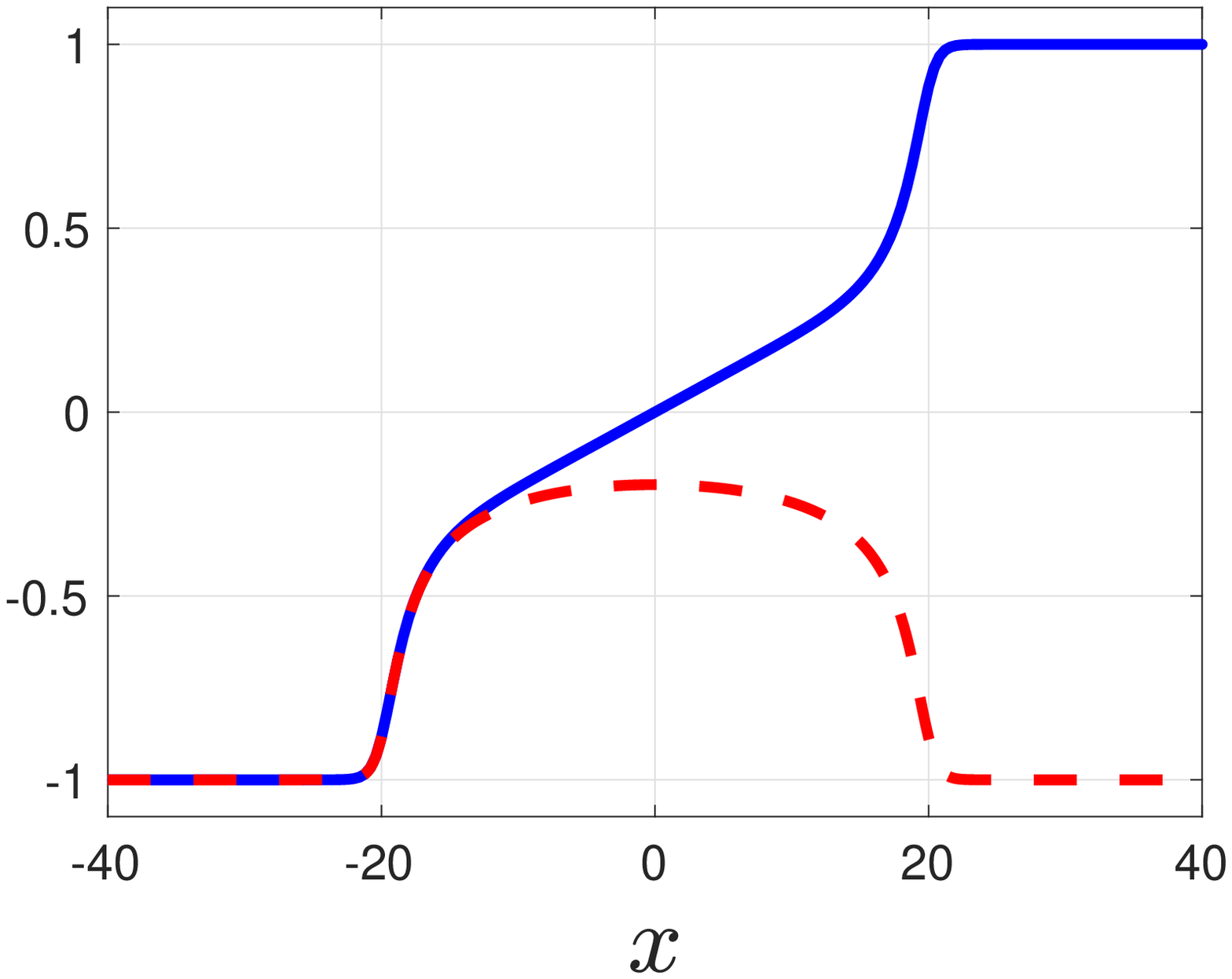}
\includegraphics[width=0.235\textwidth]{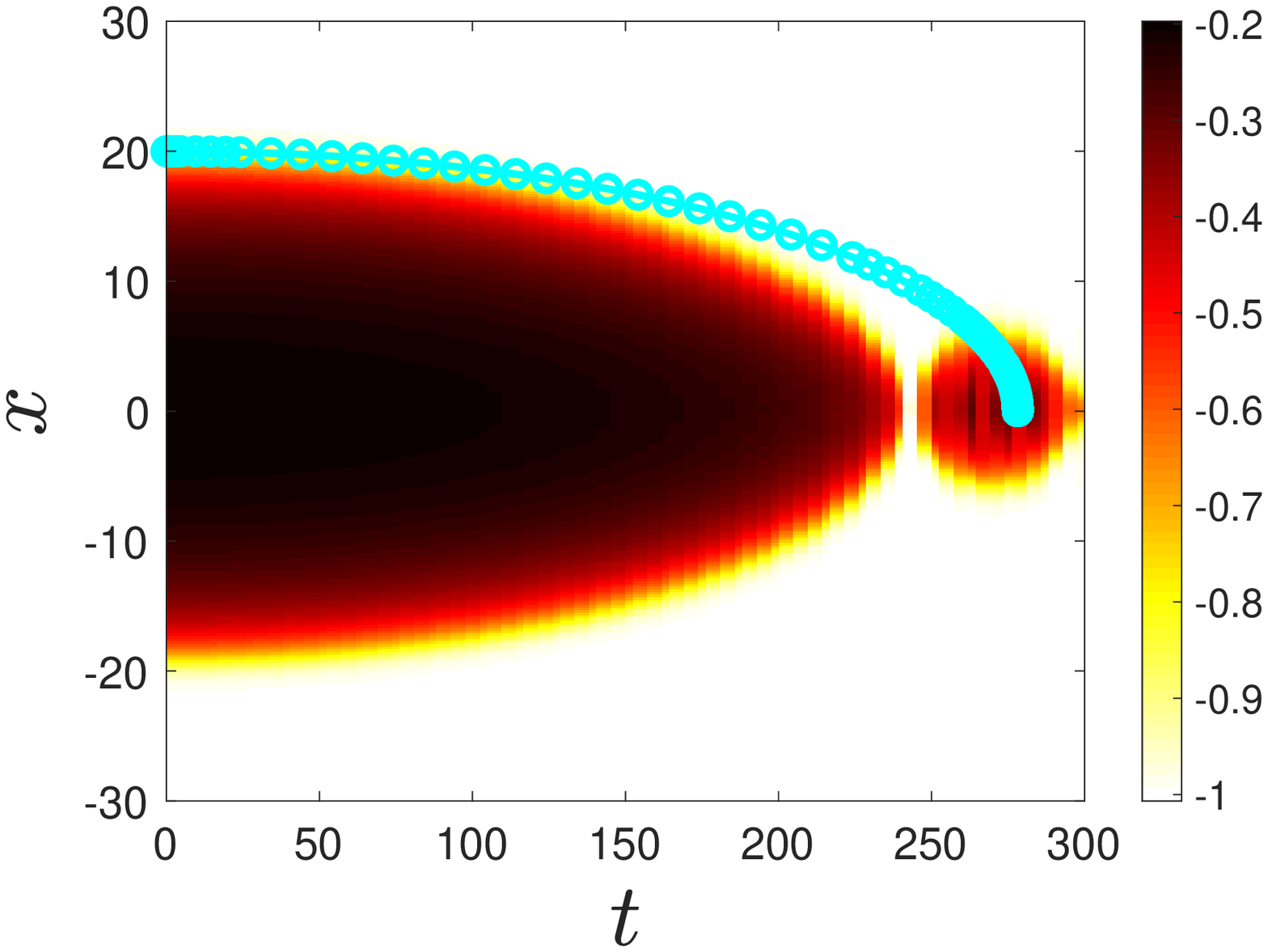}
\includegraphics[width=0.235\textwidth]{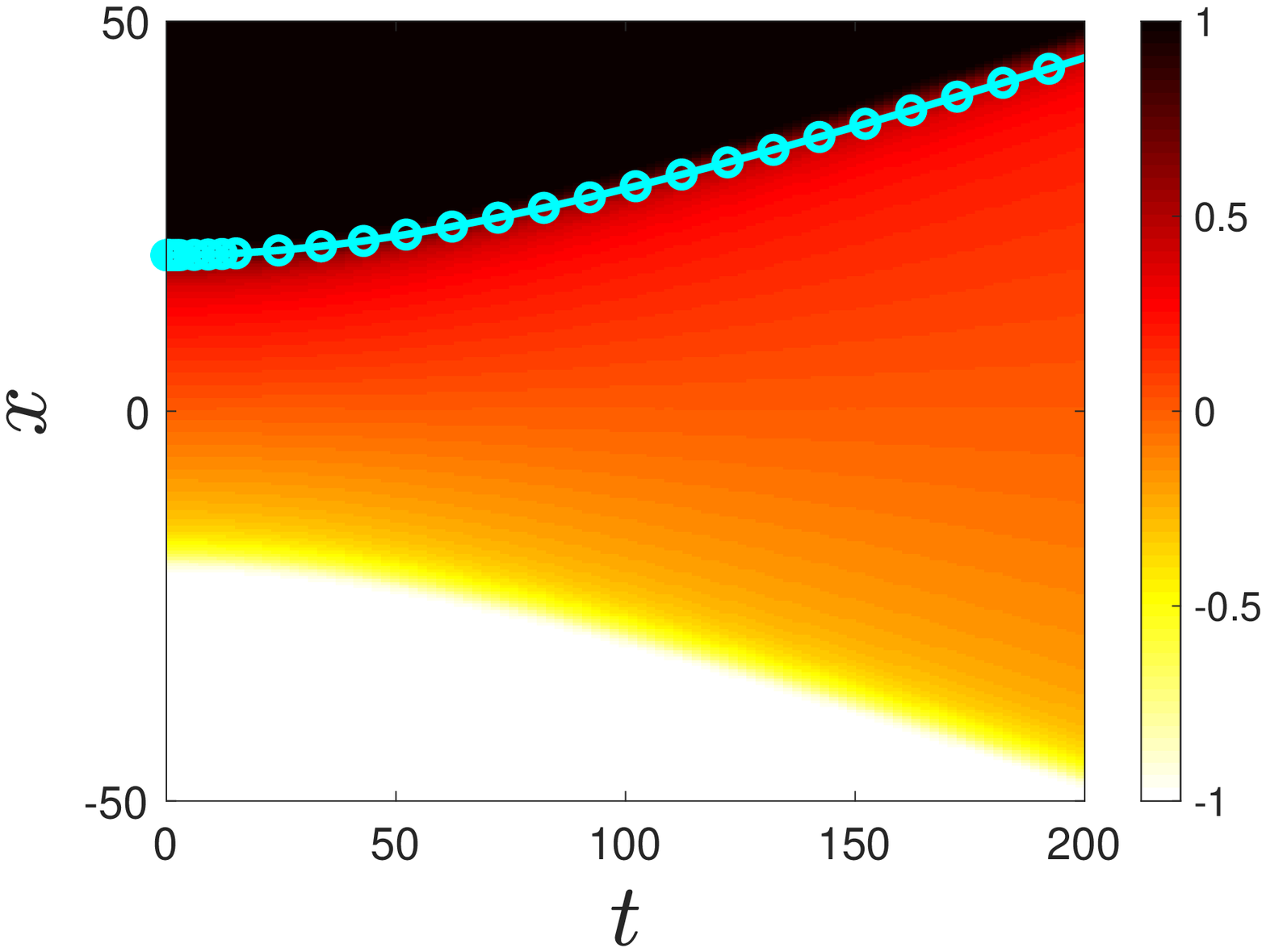}
\caption{Equivalent of Fig.~\ref{fig:powerLaw8} for the $\varphi^{10}$ model. Bottom left is space-time contour plot of the K-AK interaction, and the cyan curve with circle symbols is the solution to the IVP: $\ddot{x}(t)=-2.0676/x^3$, $x(0)=20$, $\dot{x}(0)=0$. Bottom right is space-time contour plot of the K-K interaction, and the cyan curve with circle symbols is the solution to the IVP: $\ddot{x}(t)=16.5411/x^3$, $x(0)=20$, $\dot{x}(0)=0$.}
\label{fig:powerLaw10}
\end{figure}

Here, we deploy our recent methodology~\cite{ivan_longrange}, which is critical to obtaining accurate simulations of the interactions between topological defects with power-law tails (long-range interactions). Briefly, a pseudospectral differentiation matrix with periodic boundary conditions \cite{trefethen} replaces the spatial derivatives in Eq.~(\ref{eq:nkg}) on the interval $x\in[-200,200]$ with $N=2000$ discrete points (hence, the grid spacing is $\Delta x=0.2$). The resulting system of ODEs (after discretizing in $x$) is integrated numerically using Matlab's \texttt{ode45} solver with built-in error control.

Following~\cite{ivan_longrange}, for the K-AK interactions we start with a ``split-domain'' ansatz $\varphi_{\mathrm{split}}(x)=[1-H(x)]\varphi_{(-1,0)}(x)+H(x)\varphi_{(0,-1)}(x)$, where $H(x)$ is the Heaviside unit step function. That is,  $\varphi_{\mathrm{split}}(x)=\varphi_{(-1,0)}(x)$ on the interval $(-\infty ,0]$, while  $\varphi_{\mathrm{split}}(x)=\varphi_{(0,-1)}(x)$ on the interval $(0,\infty )$. Then, $\varphi_{\mathrm{split}}(x)$ is used as the initializer for the Matlab function \texttt{lsqnonlin},  which minimizes (using nonlinear least squares) the $l_{2}$ norm of the discretized version of the opposite of the right-hand side of Eq.~(\ref{eq:nkg}), subject to the two additional constraints of keeping the positions of the kink and antikink fixed. The result is a smoothed and minimized version of $\varphi _{\mathrm{split}}(x)$, which is then used as the initial condition for solving the PDE~\eqref{eq:nkg} numerically.

The K-K case is similar, except that $\varphi_{\mathrm{split}}(x) = [1-H(x)]\varphi_{(-1,0)}(x) + H(x)\varphi_{(0,1)}(x)$. As a result, there is a discontinuity at $x=0$, which becomes large for $\varphi^{10}$ and even larger for $\varphi^{12}$. With $N=2000$, \texttt{lsqnonlin} fails to converge for some cases; however, for smaller $N$ it does converge. Thus, the output from smaller $N$ can be used as the initializer for \texttt{lsqnonlin} with $N=2000$, which then converges quickly. Except for this detail, the procedure is the same as for the K-AK case. As explained in~\cite{ivan_longrange}, this type of minimization
procedure is {\it crucial} in order to avoid inaccurate interaction observations stemming
from a more naive sum or product (of kinks) ansatz.

\begin{figure}[tbp]
\centering
\includegraphics[width=0.235\textwidth]{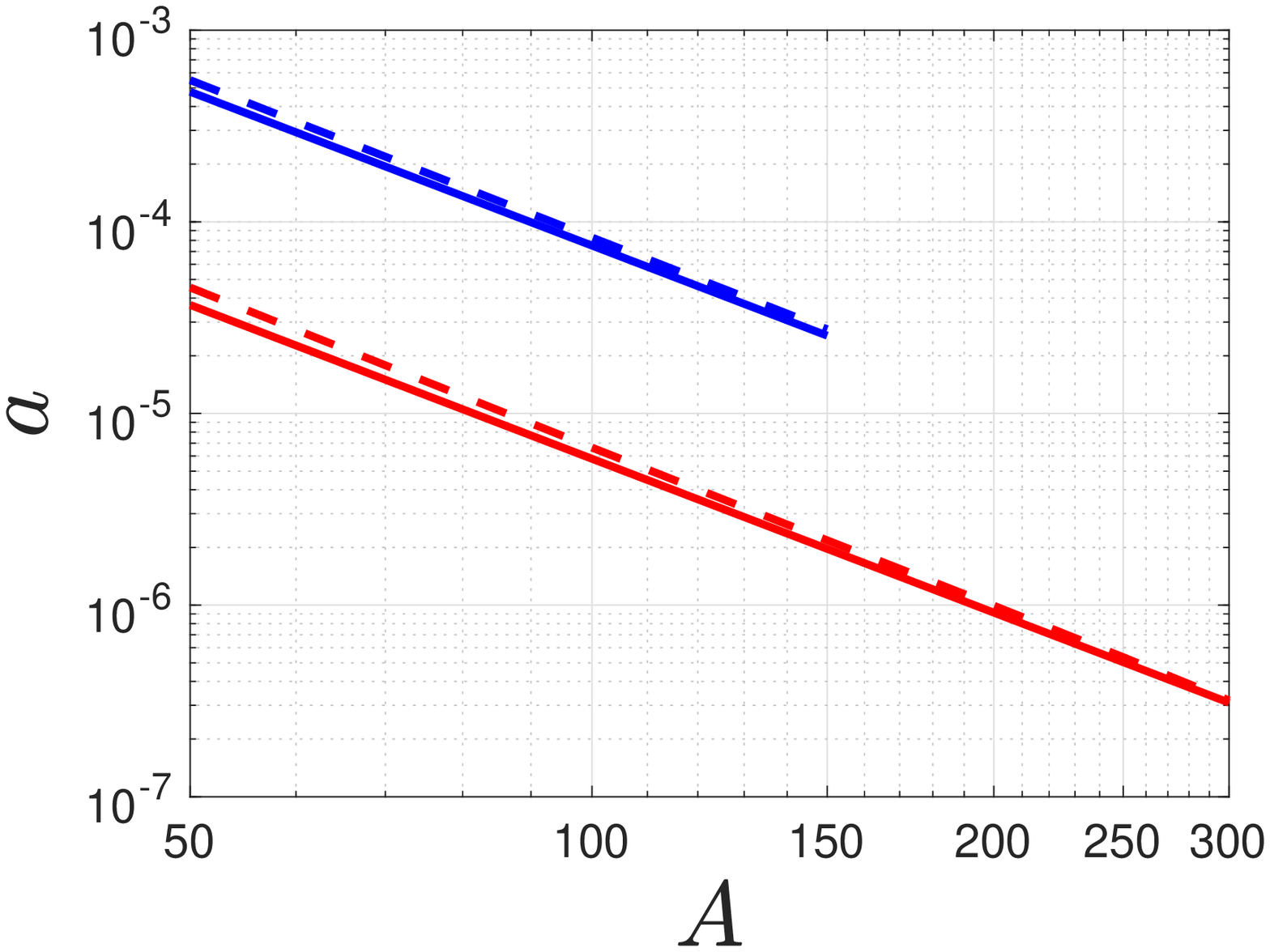}
\includegraphics[width=0.235\textwidth]{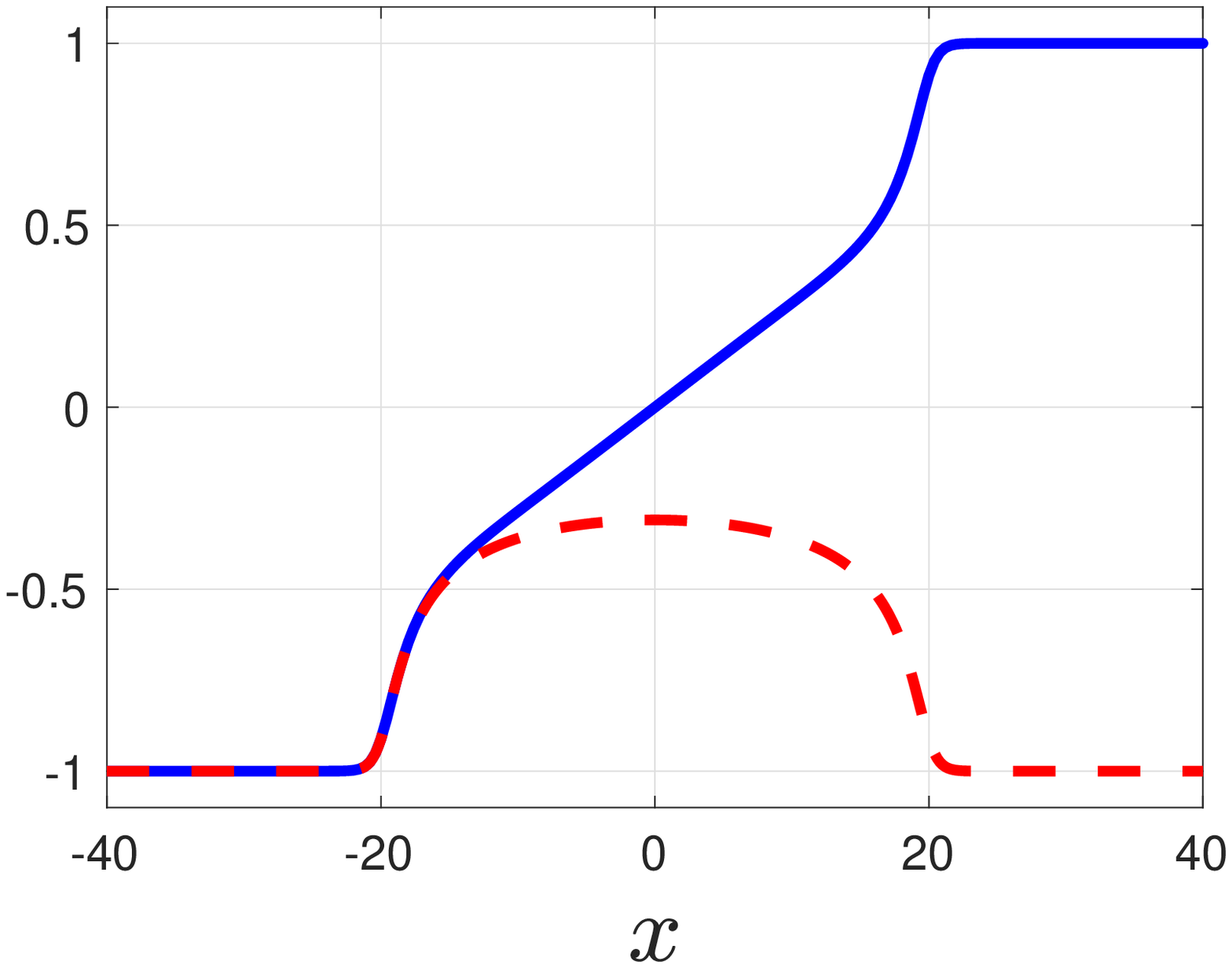}
\includegraphics[width=0.235\textwidth]{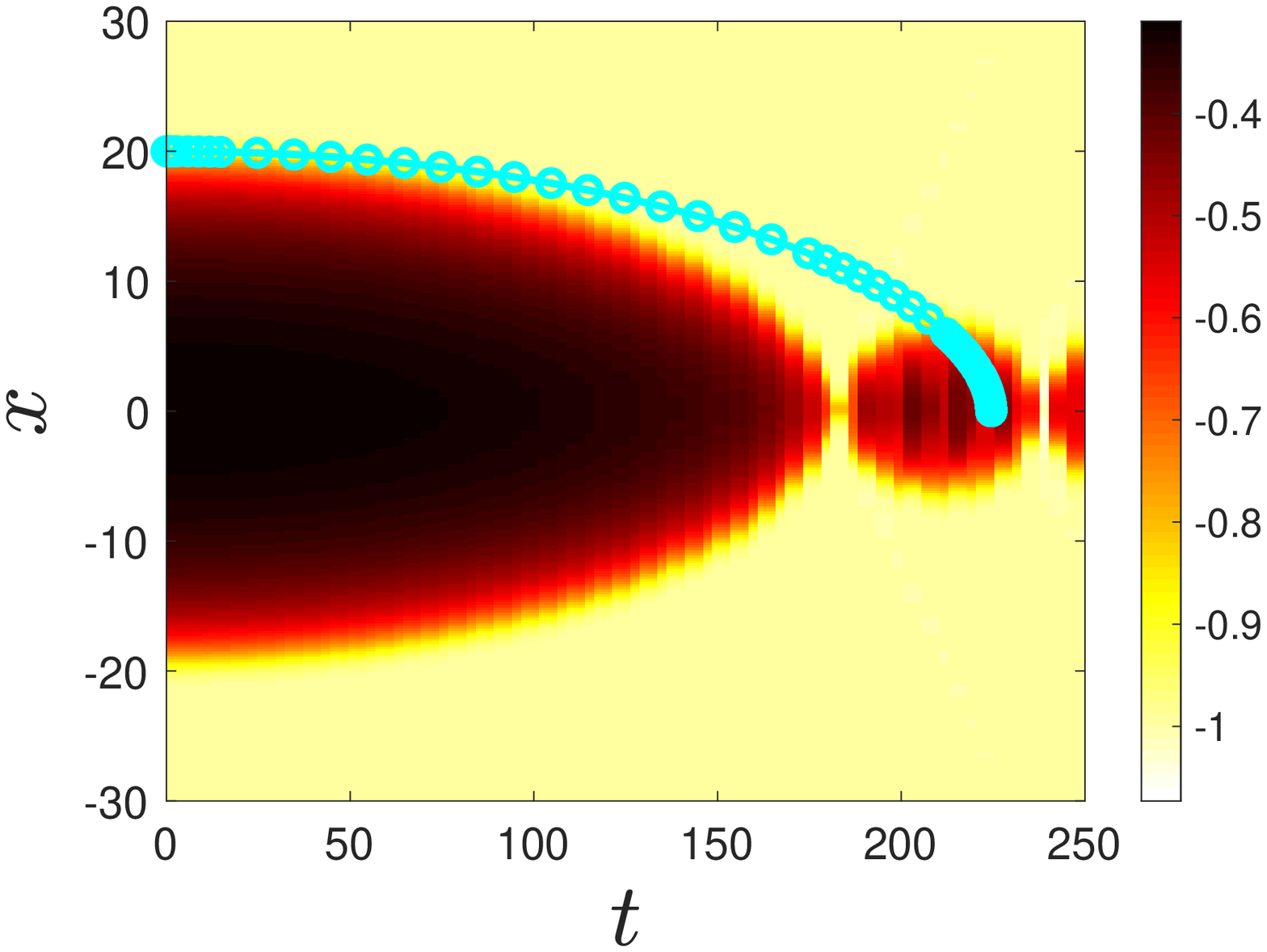}
\includegraphics[width=0.235\textwidth]{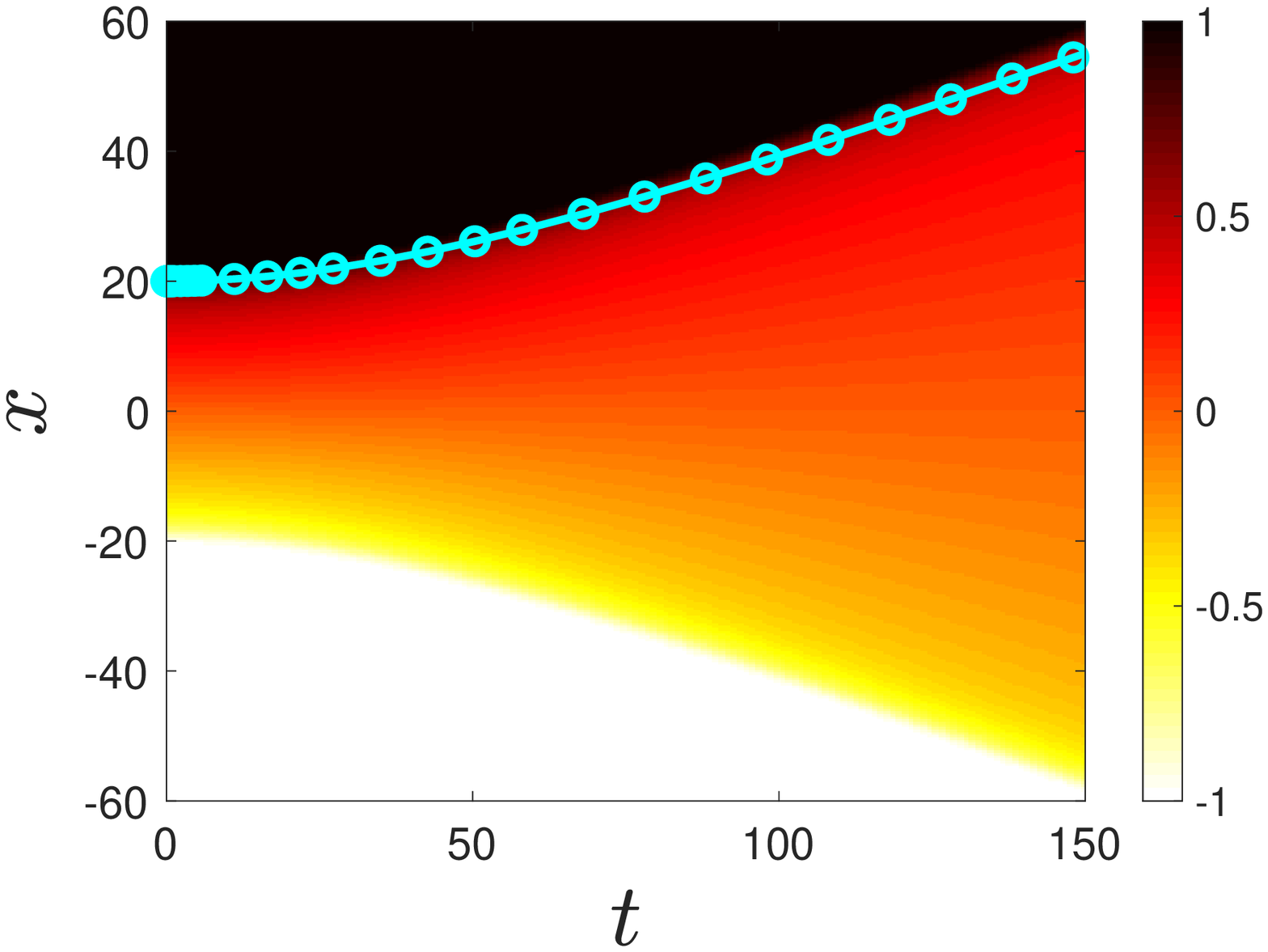}
\caption{Equivalent of Figs.~\ref{fig:powerLaw8} and \ref{fig:powerLaw10} for the $\varphi^{12}$ model. 
Bottom left is space-time plot of the K-AK interaction, and the cyan curve with circle symbols is the solution to the IVP: $\ddot{x}(t)=-1.2495/x^{8/3}$, $x(0)=20$, $\dot{x}(0)=0$. Bottom right is space-time contour plot of the K-K interaction, and the cyan curve with circle symbols is the solution to the IVP: $\ddot{x}(t)=16.1871/x^{8/3}$, $x(0)=20$, $\dot{x}(0)=0$.}
\label{fig:powerLaw12}
\end{figure}

In Figs.~\ref{fig:powerLaw8}--\ref{fig:powerLaw12}, the top-right panel shows the  K-K and K-AK configuration initializers used to evolve the PDE, i.e., Eq.~\eqref{eq:nkg}, under the $\varphi^{8}$, $\varphi^{10}$, and $\varphi^{12}$ models, respectively. The bottom-left and bottom-right panels of each figure show, respectively, the space-time evolution of the field for the K-AK case (attraction) and K-K case (repulsion). Cyan curves with circle symbols are solutions to the Newtonian equation of motion $Ma=F$, where $F$ was obtained above in the form $F=M\gamma_i(n)A^{\frac{2n}{1-n}}$  with $\gamma_i(n)$ as the corresponding prefactor ($i=\text{K-K}$ or K-AK). Since $a=\ddot{A}$ for K-K and $a=-\ddot{A}$ for K-AK, then both cases require solving the initial value problem (IVP) for the kink location $x=A$: $\ddot{x}=\pm\gamma_i(n)x^{\frac{2n}{1-n}}$, $\dot{x}(0)=0$, and $x(0)$ given.

Note that in all three figures, especially in Figs.~\ref{fig:powerLaw10} and \ref{fig:powerLaw12}, the bottom-left plots show that the attractive force between the kink and antikink leads them to collide at $x=0$ at some instant of time, upon which ``bounces'' are observed. Our theory of the interaction force is asymptotic for large separation, therefore it does not in any way address the instant of {collision} and beyond. Therefore, the cyan curves (with circle symbols) {are} not expected to agree with the contours of the numerical solution as the kink and antikink locations approach the origin ($x=0$).

Next, we numerically calculate the relation between the location of the kink  (i.e., half-separation) $A$ and its acceleration $a$ (from rest) by solving the PDE in Eq.~\eqref{eq:nkg} over a very short time interval (from $t=0$ to $t=0.01$). $A$ is then calculated as a function of $t$ over this interval, which is used to estimate the acceleration $a=\ddot{A}$, which is nearly constant during this $t$-interval. Then, a least-squares model of the form $a={b}/{A^{k}}$ was fit to the simulation data. The numerically fit results are shown graphically in log-log plots in the upper-left panels of Figs.~\ref{fig:powerLaw8}--\ref{fig:powerLaw12}. Therein, the numerically fit models are also compared to the results from the theoretical analysis above. 
The fit between the asymptotic prediction  and the numerical results is good in all the cases considered, and nearly perfect for the $\varphi^{8}$ model. The kink location predicted by solving the appropriate IVP is superimposed onto the contour plots in the bottom panels of Figs.~\ref{fig:powerLaw8}--\ref{fig:powerLaw12}.

In Tables \ref{tab:table1} and  \ref{tab:table2}  we summarize our findings, both theoretical and numerical.
In calculating the error between the theoretical and numerical models we find that, for smaller values of $A$, the error between the models is greater
(especially as $n$ becomes larger). The reason for this discrepancy is twofold: (i) the theoretical model derived above is valid only asymptotically for large separations, and (ii) for large $n$, the domain walls exhibit ``fatter'' tails, thus it becomes increasingly difficult to prepare a ``well-separated'' initial condition. Therefore, we restrict ourselves to $A\geq 50$ when calculating the fit to the numerical simulation data and when comparing it against the theoretical prediction. 

For the K-AK interaction, we used six $A$ values (data points) in the interval $[50,300]$, while for the K-K interaction we used six $A$ values in the interval $[50,150]$. For the K-K case it is difficult to find accurate initial conditions for the PDE for $A>150$ (because Matlab's \texttt{lsqnonlin} takes longer, or fails, to converge to an appropriate field configuration to be used as an initial condition). The relative error between the theoretical model and the numerical fit is calculated over the same range as the range of data points used to obtain the numerical models. In all cases, the maximum error occurs at the first data value ($A=50$). While a more computationally intensive investigation of the suitable distance regime in which the asymptotic theoretical predictions are valid may significantly reduce the error  in Tables~\ref{tab:table1} and \ref{tab:table2} for larger $n$, on the basis of the currently available results, we conclude that further investigation would be required to determine such a range.

\begin{table}[tbp]
  \begin{ruledtabular}
    \begin{tabular}{lcccc} 
      $n$ & {Theory} & {Fit} & {Range} & {Error}\\
      \hline
      $2$ &${44.31}/{A^4}$ & ${43.43}/{A^{3.996}}$ & $50 \leq A \leq 150$& $0.35\%$\\
      $3$  &${16.54}/{A^3}$ & ${21.74}/{A^{3.046}}$ & $50 \leq A \leq 150$& $10\%$\\
      $4$  &${16.19}/{A^{8/3}}$ & ${23.23}/{A^{2.724}}$& $50 \leq A \leq 150$ & $15\%$\\
    \end{tabular}
        \caption{Theoretical model and numerical fit model predictions for the acceleration as a function of position during the kink-kink interaction. The error is the maximum relative error between the theory and fit curves over the specified range of $A$.} 
    \label{tab:table1}
  \end{ruledtabular}
\end{table}

\begin{table}[tbp]
  \begin{ruledtabular}
    \begin{tabular}{lcccc} 
      $n$ & {Theory} & {Fit} & {Range} & {Error}\\
      \hline
      $2$ &${11.08}/{A^4}$ & ${10.92}/{A^{3.997}}$ & $50 \leq A \leq 300$& $0.4\%$\\
      $3$  &${2.068}/{A^3}$ & ${3}/{A^{3.064}}$ & $50 \leq A \leq 300$ & $13\%$\\
      $4$  &${1.25}/{A^{8/3}}$ & ${2.234}/{A^{2.762}}$ & $50 \leq A \leq 300$ & $23\%$\\
    \end{tabular}
        \caption{Theoretical model and numerical fit model predictions for the acceleration as a function of position during the kink-antikink interaction. The error is the maximum relative error between the theory and fit curves over the specified range of $A$.} 
    \label{tab:table2}
  \end{ruledtabular}
\end{table}

\paragraph*{Conclusions and Future Work.} 
In the present work, we have taken a significant step beyond the standard field-theoretic models for topological defects and their interactions, which have been studied for a number of decades. Up to now, the vast majority of the associated one-dimensional efforts have focused on kinks with exponential tail decay, thus endowing the coherent structures with a ``short-range'' exponential tail-tail interaction. Using potentials with the highest power going as $\varphi^{2n+4}$, for arbitrary $n$, as the vehicle of choice in this work, we have systematically examined the long-range pairwise kink-kink and kink-antikink interactions. We have blended the state-of-the-art asymptotic tools with carefully crafted numerical simulations to elucidate the power-law nature of the decay of the interaction force with the $(\frac{2n}{n-1})$th power of the separation between the topological defects. Equally important, we have identified the prefactor of this interaction (for arbitrary $n$) and have confirmed its agreement with numerical simulations for $n=2$, $n=3$, and $n=4$.

Our results will likely provide valuable insights into domain wall interaction in materials \cite{ferro, iso} and biophysical \cite{proteins} contexts that are governed by higher-order field theories. We also hope that this study will pave the way for the formulation of novel collective coordinate treatments \cite{weigel02} of long-range interactions, and a systematic understanding of their outcomes (including the role of initial kinetic energy; here, to crystallize the relevant phenomenology we restricted ourselves to kinks initially at rest). Another direction of future work concerns the exploration of coherent structures in higher dimensions \cite{greene15} and the understanding of the existence, stability and dynamics of localized and vortical patterns therein. Finally, the methodology developed herein can be applied to kink interactions in other recently proposed higher-order field theories harboring power-law tails \cite{khare,bazeia18,KS}.

\paragraph*{Acknowledgments.} This material is based upon work supported by the U.S.\ National Science Foundation under Grant No.\ PHY-1602994 and under Grant No.\ DMS-1809074 (P.G.K.). The work of MEPhI group was supported by the MEPhI Academic Excellence Project under Contract No.\ 02.a03.21.0005. V.A.G.\ also acknowledges the support of the Russian Foundation for Basic Research under Grant No.\ 19-02-00971. A.K.\ is grateful to INSA (Indian National Science Academy) for the award of INSA Senior Scientist position.  A.S.\ was supported by the U.S.\ Department of Energy.

\end{document}